\documentclass[namedreferences]{kluwer}

\usepackage{epsfig}
\newcommand{\farcs}{\hbox{$.\!\!^{\prime\prime}$}}

\begin{document}
\begin{article}
\begin{opening}

\title{What will the next generation radio telescope detect at 1.4\,GHz?}

\author{A. \surname{Hopkins}\email{ahopkins@atnf.csiro.au}}
\institute{Australia Telescope National Facility, PO Box 76 Epping, NSW 1710,
  Australia.}
\author{R. \surname{Windhorst}}
\institute{Dept. of Physics and Astronomy, Arizona State University, Box 871504,
  Tempe, AZ 85287-1504, USA}
\author{L. \surname{Cram}}
\institute{Astrophysics Department, School of Physics, University of Sydney,
  NSW 2006, Australia}
\author{R. \surname{Ekers}}
\institute{Australia Telescope National Facility, PO Box 76 Epping, NSW 1710,
  Australia.}

\runningtitle{Sky simulations at 1.4\,GHz}
\runningauthor{Hopkins et al.}

\begin{abstract}
An international project is underway to design and build a radio
telescope with an effective collecting area two orders of magnitude greater
than the largest existing instruments. One of the many scientific goals of this
instrument will be the investigation of the extragalactic radio source
population at flux densities two to three orders of magnitude fainter than
the limits of existing observations. We present simulations of the radio
sky at 1.4\,GHz down to a flux density limit of $0.1\,\mu$Jy using
extrapolations of known radio luminosity functions for two different
population scenarios. The resulting simulations confirm that a resolution
of $0\farcs1$ is necessary to avoid formal confusion, but source
blending may still dominate if the intrinsic size of such faint sources
is larger than a few kiloparsecs.
\end{abstract}


\end{opening}

\section{Introduction}

New developments in all fields of astronomy have brought the current generation
of astronomers to the brink of probing the origin and evolution of the Universe
as a whole. Many new instruments, both ground- and space-based, are being
designed and built to facilitate these studies, and it has long been
recognised that major scientific advances follow such technical innovation
(\opencite{Har:81}; \opencite{dSP:63}, for example). To maintain the
extraordinary
momentum of discovery of the last few decades in the metre and centimetre
wavelength regime, a very large new radio telescope will be needed, having
a sensitivity 100 times better than existing premier radio telescopes.
An increase in sensitivity of this order cannot be achieved by improving the
electronics of receiver systems, but only by having a radio telescope
with a total effective collecting area of about a million square metres.
The instrument has therefore acquired the appellation the ``Square
Kilometre Array" (SKA). The time by which such a new international world radio
observatory is needed to complement other planned instruments will be in the
years around 2010. The international effort to design and build the SKA
has been growing and progressing since 1993, when the International Union for
Radio Science (URSI) established the Large Telescope Working Group to begin
developing scientific goals and technical specifications for the next
generation radio observatory.

A more detailed introduction to the SKA project, including preliminary
specifications, is given by \inlinecite{BT:99}, who outline the SKA
concept and present a wide range of the scientific questions such an
instrument could address. These include a number of key goals, such as
establishing the redshift of reionisation (\opencite{Shav:99}, for example),
detection of the precursors of the first galaxies through H{\sc I},
measuring the mass function of black holes in galaxy cores,
pulsar detection and timing in other galaxies as well as our own,
and
investigation of star-formation processes and radio emission from stars,
among many others no less inspiring \cite{BT:99,Hop:99c}. In addition, there
will be unanticipated discoveries that may represent the most significant
results from the instrument. Since we cannot predict the nature of
serendipitous discoveries, however, we are limited to investigating
identified goals. This paper investigates the implications of our current
knowledge about extragalactic radio source populations on the results of
anticipated 1.4\,GHz radio continuum observations with such an instrument.
This frequency has been chosen since it is one of the most well studied,
with radio source counts measured over almost seven orders of magnitude.

A number of published radio surveys probe the population of microjansky
($<0.1\,$mJy) radio sources
\cite{Wind:93a,Wind:95,Rich:98,Hop:98,Hop:99b,Nor:99}.
Below a few mJy, the 1.4\,GHz population begins to be dominated by
starburst and disk galaxies, whose radio emission derives primarily from star
formation processes \cite{Wind:85,Kro:85,Ben:93,Fom:97,Age:99}, as opposed to
the classical radio sources fuelled by active galactic nuclei (AGN), which
dominate at higher flux densities. Interest in the sub-millijansky
population stems from the desire to investigate the nature and history
of star-formation in galaxies, and the potential exists
through radio-selection to provide very large, homogeneously
selected samples of actively star-forming galaxies over a wide range in
redshift. While current surveys are only just starting to compile large
samples, approaching thousands of objects \cite{Age:99,Hop:99b}, the
anticipated sensitivity at nJy levels, resolution of $0\farcs1$, and field size
of one square degree at 1.4\,GHz, imply that surveys with the SKA
will rapidly dwarf present surveys.
To investigate the issues surrounding source confusion and dynamic range
limitations in such surveys, we use extrapolations of current models
describing the 1.4\,GHz radio population, and have begun a program of
simulating aspects of the radio sky at this wavelength.
Throughout this paper we use $H_0=50\,$km\,s$^{-1}$\,Mpc$^{-1}$ and $q_0=0.5$.

\section{The model}

The local 1.4\,GHz luminosity function (LF) is well determined down
to $L \approx 10^{18.5}\,$WHz$^{-1}$ \cite{Con:89}. LFs invoking luminosity
evolution models for the AGN population \cite{DP:90} and the starburst
population \cite{Row:93,Hop:98} match the observed 1.4\,GHz source counts
down to the flux density limits of current observation, and we
have adopted these models for our investigation. We have extrapolated these 
1.4\,GHz LFs to luminosities much fainter than have been observed, and
supplemented them with distributions in linear size of the objects. This
effectively provides a ``model universe," which is then projected onto the
desired two-dimensional field of view. Although the effects of large-scale
structure have been ignored in this initial model, they will be included
as the work is refined, adding the ability to test various
clustering models against observation. The model also allows
the simulation of different cosmologies, and different rates and types of
evolution for direct comparison to observation. In practice, though, this
requires the development of different luminosity function models and
evolutionary rates for each cosmology of interest, to ensure the
source counts predicted from the model are consistent with existing
observations.

\subsection{Radio source populations}

The galaxy populations used for our simulations comprise four
distinct types of radio sources that we broadly classify as ``AGN" or
``starforming" according to whether the radio emission for the population
is derived predominantly from a central engine or from star-formation
processes, respectively. We have divided the AGN class into steep-spectrum and
flat-spectrum radio sources, following \inlinecite{DP:90}.
The LFs corresponding to these two populations have the form
\cite{DP:90}:
\begin{equation}
\phi = \phi_0\left\{\left(\frac{L_{1.4}}{L_c(z)}\right)^{\alpha} + 
\left(\frac{L_{1.4}}{L_c(z)}\right)^{\beta}\right\}^{-1},
\end{equation}
with parameters
\begin{eqnarray*}
\phi_0 & = & 10^{-6.91}, \\
\alpha & = & 0.69, \\
\beta  & = & 2.17, \\
L_c(z) & = & 26.22 + 1.26z - 0.26z^2,
\end{eqnarray*}
for steep spectrum AGN, and
\begin{eqnarray*}
\phi_0 & = & 10^{-8.15}, \\
\alpha & = & 0.83, \\
\beta  & = & 1.96, \\
L_c(z) & = & 26.36 + 1.18z - 0.28z^2,
\end{eqnarray*}
for flat spectrum AGN. The luminosity evolution of this population is included
in the redshift dependence of $L_c(z)$. This evolution results in a redshift
cutoff in the population beyond $z\approx2$ \cite{DP:90}, which can be seen
in the redshift distributions from the simulations (Section~\ref{simul}).

The starforming class includes what we refer to as ``IRAS-type" galaxies,
for which we have used the $60\,\mu$m ``warm IRAS"
galaxy luminosity function of \inlinecite{Sau:90}, after conversion to
1.4\,GHz assuming $L_{60}=100L_{1.4}$ \cite{Row:93}. This IRAS-type
population is identified with the ``starburst" galaxy population that begins
to dominate the 1.4\,GHz source counts below a few millijanskies.
Their luminosity function has the form \cite{Sau:90,Row:93}:
\begin{equation}
\phi(L) = C \left(\frac{L}{L^*}\right)^{1-\alpha}
\exp\left[-\frac{1}{2\sigma^2} \log^2_{10}\left(1 +
\frac{L}{L^*}\right)\right],
\end{equation}
with parameters (assumed the same at $60\,\mu$m and 1.4\,GHz) of
$\alpha=1.27$, $\sigma=0.626$, and $C=3.25\times10^{-2}$. At $60\,\mu$m,
$L^*_{60}=10^{23.9}$\,WHz$^{-1}$ so from the radio/FIR correlation,
$L^*_{1.4}=10^{21.9}$\,WHz$^{-1}$. The luminosity evolution for
this population is of the form $L(z)\propto(1+z)^Q$ with $Q=3.3$. A form
of redshift cutoff at $z=2$ was used in this evolution, also. For $z>2$ the
evolution was calculated as though the source was at $z=2$, i.e.
$L(z\ge2)\propto(1+2)^Q$ \cite{Hop:98}.

Radio emission produced through star-formation processes can be
interpreted as an indicator of the current star-formation rate (SFR) in a
galaxy if no AGN emission is present \cite{Cram:98}. Hence, as observational
sensitivity improves, the so-called ``normal" galaxies will eventually be
detectable through their radio emission even though their SFRs may be quite
low. The population of normal galaxies so detected may be adequately
represented by the faint end tail of the ``warm
IRAS" luminosity function, and we have constructed several simulations under
this assumption. There is the alternative possibility that an additional
luminosity function needs to be invoked to account for galaxies with only
modest levels of star formation. Certainly the radio source counts derived
only from the ``warm IRAS" luminosity function (combined with the AGN
population) lie well below the upper limit derived from limits on
the cosmic background radiation (CBR) fluctuations
\cite{Fom:93,Wind:93b,Wind:99}.

\begin{figure}
\centerline{\epsfig{figure=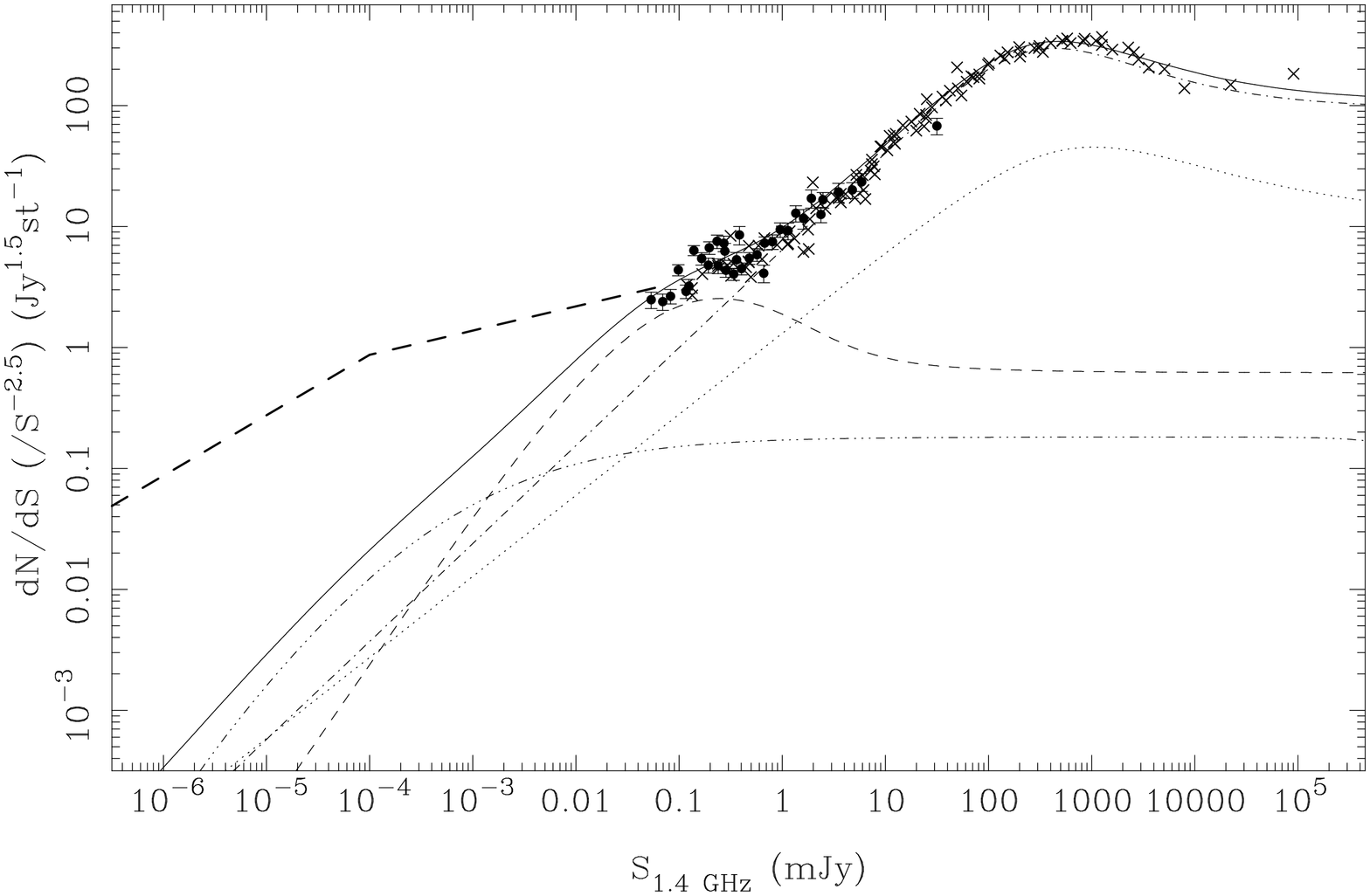,width=5cm,angle=-90}}
\caption{Models of the source counts showing relative contributions of
different populations. Thin dashed line: IRAS-type population; Dotted
line: AGN flat-spectrum population; Dot-dashed line: AGN steep-spectrum
population; Dot-dash-dotted line: normal galaxy population; Solid line: sum
of all populations. The symbols are existing observational source
counts (see text). The normal galaxy population shown here is the maximum
possible contribution supplementary to the other models and consistent with
the observations. The thick dashed line represents the upper limit to the
source counts implied from CBR constraints. More details are given in the
text.}
\label{scmod}
\end{figure}

To include these modest starformers in our simulations, we have chosen to
use the known luminosity function for optical galaxies, the great majority
of which have low SFRs. Accurately modelling the radio properties from
the optical LF, however, requires a knowledge of the
bivariate (radio-optical) luminosity function (BLF) over a large range of
luminosities. Despite the limits to the current knowledge of the BLF
\cite{Aur:77,Sad:89,Hop:99b}, an attempt to include this population was
made using an over-simplified model: the assumption of a constant radio/optical
luminosity ratio. One possibility for refining this assumption
could be to use a luminosity ratio distribution closer to the observed
distribution in the far-infrared/optical luminosity ratio
\cite{Cor:91,Row:87}, by invoking the radio/FIR correlation, although
we have not explored this extra parameter space in the current investigation.
The optical ($B_J$-band) luminosity function used \cite{Efs:88} was a
Schechter function with parameters $\alpha=-1.1$, $\phi^*=0.0156$ and
$M^*=-19.9\,$mag, which corresponds to $L^*=10^{21.5}\,$WHz$^{-1}$ at 440\,nm.
The luminosity ratio chosen was $L_{1.4}/L_{opt}=\frac{1}{3}$ (for
$L_{1.4}$ and $L_{opt}$ in units of WHz$^{-1}$), which corresponds to a
SFR of $\approx0.26\,M_{\odot}$yr$^{-1}$ for an optical $L^*$ galaxy
(from equation~1 of \opencite{Cram:98}). This ratio is the highest possible
(constant) value consistent with the observed counts and the contributions
from the other populations. The resulting luminosity function (with no
evolution invoked), combined with
the other populations, has been used to predict the source counts to a
flux density of 1\,nJy (Figure~\ref{scmod}). The symbols in this Figure
come from two main sources, the crosses are a compilation of 1.4\,GHz source
counts from \inlinecite{Wind:93a}, and the circles from the
{\em Phoenix Deep Survey\/} (\opencite{Hop:99b}, and references therein).
The thick dashed line shown in this and succeeding Figures is the upper
limit to the source counts derived from known limits to the CBR fluctuations
\cite{Fom:93,Wind:93b,Wind:99}. If the source count continues toward fainter
flux densities from the limits of existing observations with the constant
slope seen between about 0.05--1\,mJy ($\gamma=2.3$, where the differential
source counts $n(S) \propto S^{-\gamma}$), then the count must ultimately
converge below about 100\,nJy ($\pm0.5$ dex) with a slope $\gamma \le 2$.

\begin{figure}
\centerline{\epsfig{figure=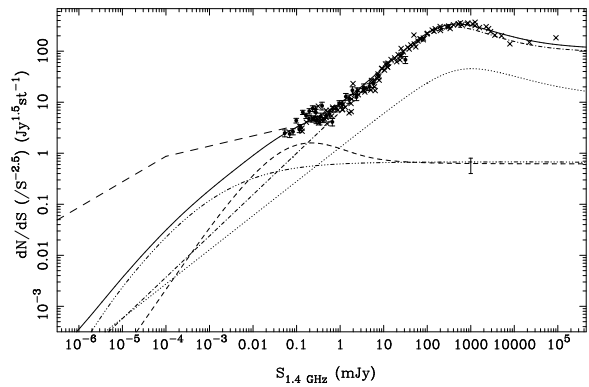,width=5cm,angle=-90}}
\caption{Models of the source counts. Here the evolution of the IRAS-type
population is reduced from $Q=3.3$ to $Q=2.9$ and a greater contribution
is made by the normal galaxies. The error bar shown on the IRAS-type source
count indicates the range permitted by the uncertainty in the luminosity
function normalisation. See text for details. Symbols and line
styles are as for Figure~\protect\ref{scmod}.
The upper limit to the source counts is again shown as the thick dashed line.}
\label{scmod2}
\end{figure}

It is possible that there could be a greater contribution from the normal
galaxy population if the rate of luminosity evolution invoked for the IRAS-type
galaxies were lower. Such a scenario is reflected in the source counts shown 
in Figure~\ref{scmod2}. Here we show source counts predicted
under the assumption that the IRAS-type galaxies undergo luminosity evolution
at a rate $L(z)\propto(1+z)^Q$ with $Q=2.9$ rather than the $Q=3.3$ rate
used in Figure~\ref{scmod}. The luminosity ratio used for the normal
galaxies in Figure~\ref{scmod2} is $L_{1.4}/L_{opt}=0.8$ (corresponding
to a SFR of $\approx0.63\,M_{\odot}$yr$^{-1}$ for an optical $L^*$ galaxy).
Uncertainty in the $60\,\mu$m luminosity function normalisation leads to an
uncertainty in the magnitude of the predicted source counts. This is indicated
by the vertical error bar shown on the counts for the IRAS-type galaxies.
Any additional population should not increase the contribution of the
starforming galaxies outside this range.

The faintest counts due to the starforming populations could alternatively
be {\em even lower\/} than predicted here if supernovae in low-mass galaxies
are efficient at removing gas (\opencite{Bab:92},
but see also \opencite{Mac:98}).
In any case it is worth remembering that there are still many parameters that
are not yet observationally constrained, and we have restricted this
investigation to predictions from only two scenarios. These scenarios both
include the two components of the AGN population, combined with a starforming
population being composed of (1) IRAS-type galaxies only, and (2) IRAS-type
galaxies plus normal galaxies with $L_{1.4}/L_{opt}=\frac{1}{3}$.

\subsection{Linear source sizes}

The distribution in linear size of radio sources at 1.4\,GHz has been
studied extensively for the population we are referring to as AGN. Indeed,
the variation in angular size of radio sources with redshift is a classic
test for investigating cosmological models (\opencite{HLR:78}, for example). In
more recent work describing the variation of linear size with redshift
\cite[for example]{Gop:91,Kap:89,Onu:89,Sin:88,Oor:87,All:84} the
general consensus is that for $z\lsim1.5$ the linear size of these
(steep-spectrum) radio sources evolves as 
\begin{equation}
l = \frac{l_0}{(1+z)^3} P^{0.3},
\end{equation}
(see also \opencite{SS:90}). Here $l$ is the proper length of the
galaxy. This relation reflects the evolution of the characteristic length
($l_0$) of the steep-spectrum AGN population, combined with the distribution
(dependent on $P$) of individual source sizes within that population. A value
of $l_0 = 3\times10^{-6}\,$kpc\,(WHz$^{-1}$)$^{-0.3}$ was used for consistency
with \inlinecite{SS:90}. In our simulations we have adopted
this relationship for $z<1.5$. For larger redshifts, $z=1.5$ was used
when calculating the intrinsic linear size of a source (i.e.
$l = l_0/(1+1.5)^3 P^{0.3}$). The angular size for each source is derived
from its intrinsic linear size, $l$, in Mpc, and its actual redshift, $z$, by
\begin{equation}
\theta = l \frac{(1+z)^2}{d_{L}},
\end{equation}
where $d_{L}$ is the luminosity distance
\begin{equation}
d_{L} = \frac{c}{H_0q_0^2} (q_0z + (q_0 - 1)(\sqrt{1+2q_0z}-1)).
\end{equation}
Some recent work \cite{Blun:99} suggests that the above
para\-metrisation of linear size is too simplistic and is biased by catalogue
specific selection effects, although these conclusions are more applicable
at frequencies lower than 1.4\,GHz. For this reason we neglect the
results of \inlinecite{Blun:99} in our present simulations, and this
should be kept in mind when they are presented in Section~\ref{simul}.

There is evidence that, in galaxies with higher levels of star-formation
activity, this activity is concentrated in smaller regions than for less
active galaxies \cite{DBJ:91}. For the more modest starforming galaxies the
extent of the radio emission (coming predominantly from supernovae and
H{\sc II} regions) is likely to be larger, but primarily restricted
to the size of the optical disk. For the IRAS-type population a distribution
of 1.4\,GHz linear sizes was derived from the radio sizes of IRAS galaxies
\cite{DBJ:91,Eal:88}, and a distribution of half-light radii for disk
galaxies \cite{Roc:98} was used for the normal galaxy population. These
distributions are both luminosity dependent, and are quite different to those
used above for the steep-spectrum AGN. The relation for IRAS-types
was drawn from Figure 4 of \inlinecite{DBJ:91} and converted to 1.4\,GHz
using a spectral index of $-0.7$ giving:
\begin{equation}
\log(l) = 0.5\log(L_{1.4}) - 10.8
\end{equation}
where $l$ is the linear size in kpc and $L_{1.4}$ the luminosity in WHz$^{-1}$.
This relation was also used for the flat-spectrum AGN population, which is
typically composed of quite compact sources.
The relation for the normal galaxies is
\begin{equation}
\log(r_{\rm hl_0}) = -0.2M_B - 3.30
\end{equation}
and using an evolution of the half-light radius, $r_{\rm hl_0}$, of the form
$r_{\rm hl}= (1-0.2z)r_{\rm hl_0}$ approximated from Figure 2 of
\inlinecite{Roc:98}. This evolution was truncated at $z=4$ so that beyond this
limit $r_{\rm hl}$ was calculated as though the source was at $z=4$. These
parametrisations are intended to be representative rather
than exhaustive, as obviously galaxies of the same luminosity can have
quite different physical sizes. The values determined for the IRAS-type
galaxies are typically an order of magnitude smaller than those for the normal
population, consistent with the result found by \inlinecite{Eal:88} that radio
sizes of high FIR luminosity IRAS sources are about a tenth the size of the
optical galaxy. This is also consistent with the result that in disk galaxies
the scale-length of the disk is about ten times the effective radius
of the bulge (\opencite{deJ:96}, Figure 18, particularly K-band results,
which are less likely to be affected by extinction).

The resulting simulations are quite sensitive to the
distributions chosen here, as the numbers of starforming galaxies are very
large when faint flux density limits are chosen. Accordingly, this aspect of
the simulation would benefit from a more accurate parametrisation of the linear
sizes of faint optical galaxies as a function of luminosity and redshift.

\subsection{Simulated Images}

To construct simulated images, the luminosity functions and evolutionary models
are used to define the number, luminosity and redshift of sources in a
specified field of view, above a given flux density limit. The luminosities
are combined with redshifts and the chosen cosmology to produce flux densities
for each source. Similarly, the intrinsic sizes are used to derive apparent
angular sizes for each source. Each source is represented in the image by a
two-dimensional gaussian distribution of the total flux density, with random
position angle. The apparent ellipticity distribution of faint field galaxies
changes very little from $B=15$ to $B=27$ mag \cite{Ode:97}, so in a sense the
distribution of ellipticities is intrinsically random and hardly affected by
evolutionary effects, even though the galaxy morphology changes noticeably
between $B=15$ and $B=27$. To model the observed axial ratio, $R$, of the
gaussian for disk systems we refer to the results of \inlinecite{Lam:92}
who find that a pure oblate model for disk galaxies fails to
reproduce the observed axial ratios. However, a nearly oblate model with
intrinsic axial ratio $r=0.2$ can produce consistent fits to the
observations. \inlinecite{BV:81} also find a range $0.15<r<0.35$
derived for spiral galaxies from the Second Reference Catalogue of Bright
Galaxies \cite{dV:76}, and state that later type spirals have distributions
that peak at smaller values of $r$. For simplicity, then, we have assumed each
disk is an oblate ellipsoid with intrinsic axial ratio of $r=0.2$.
A random angle of inclination, $i$, is chosen which then gives the observed
axial ratio, $R$, as
\begin{equation}
R = \sqrt{r^2 + (1-r^2)\cos^2i}.
\end{equation}
The source so constructed is randomly positioned within the specified field of
view, and coloured according to type, red for AGNs and blue for starformers.
If sources overlap, their flux densities are simply added together,
producing appropriate colour combinations if they are not the same colour
source. No attempt to include the effects of obscuration or gravitational
lensing has been made, although these could in principle be
added to the model.

In addition, as an elementary step toward modelling the double-lobed nature
of many real AGNs, a pair of adjacent elliptical gaussians have been used for
the steep spectrum sources, rather than the single elliptical gaussian used
for flat spectrum AGNs and starformers. This is an unrealistic
over-simplification, particularly as no orientation effects, beaming, or
more complicated morphologies have been contemplated. The usefulness of
refinements to this aspect of the display can be
debated, but for the current purpose this simple step was deemed sufficient.

\section{Simulations}
\label{simul}

The algorithm described above has been used to create simulated
radio images for a wide variety of field sizes and flux density limits. Many
simulations were investigated but here we present only two,
indicative of our results, for analysis in terms of the properties of the SKA.

\begin{figure}
\centerline{\epsfig{figure=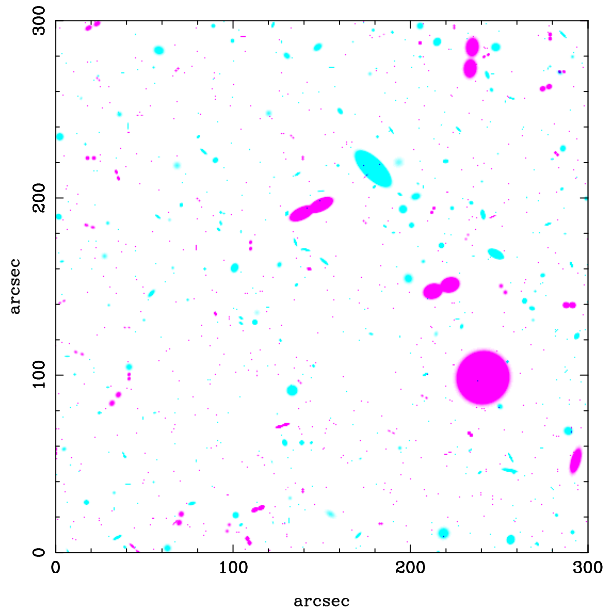,width=8cm,angle=-90}}
\caption{Simulated 1.4\,GHz image. The image size is $5'$, with a flux density
limit of $0.1\,\mu$Jy. There are 1008 objects in the simulation, divided into
two categories: AGN (red, 574 objects) and starformers (blue, 434 objects).}
\label{pics1}
\end{figure}

\begin{figure}
\centerline{\epsfig{figure=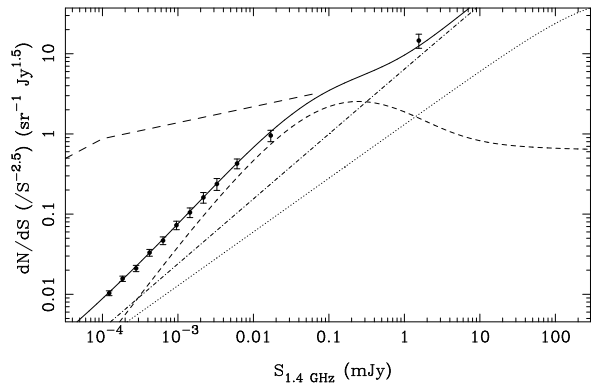,width=5cm,angle=-90}}
\caption{The source counts from the model luminosity functions
(line styles as in Figure~\protect\ref{scmod}) used to produce the image
in Figure~\protect\ref{pics1}, along with those calculated
from a typical simulation (filled circles showing Poisson error bars),
shown to confirm the self-consistency of the simulations.
The upper limit to the source counts is again shown as the thick dashed line.}
\label{sc1}
\end{figure}

\begin{figure}
\centerline{\epsfig{figure=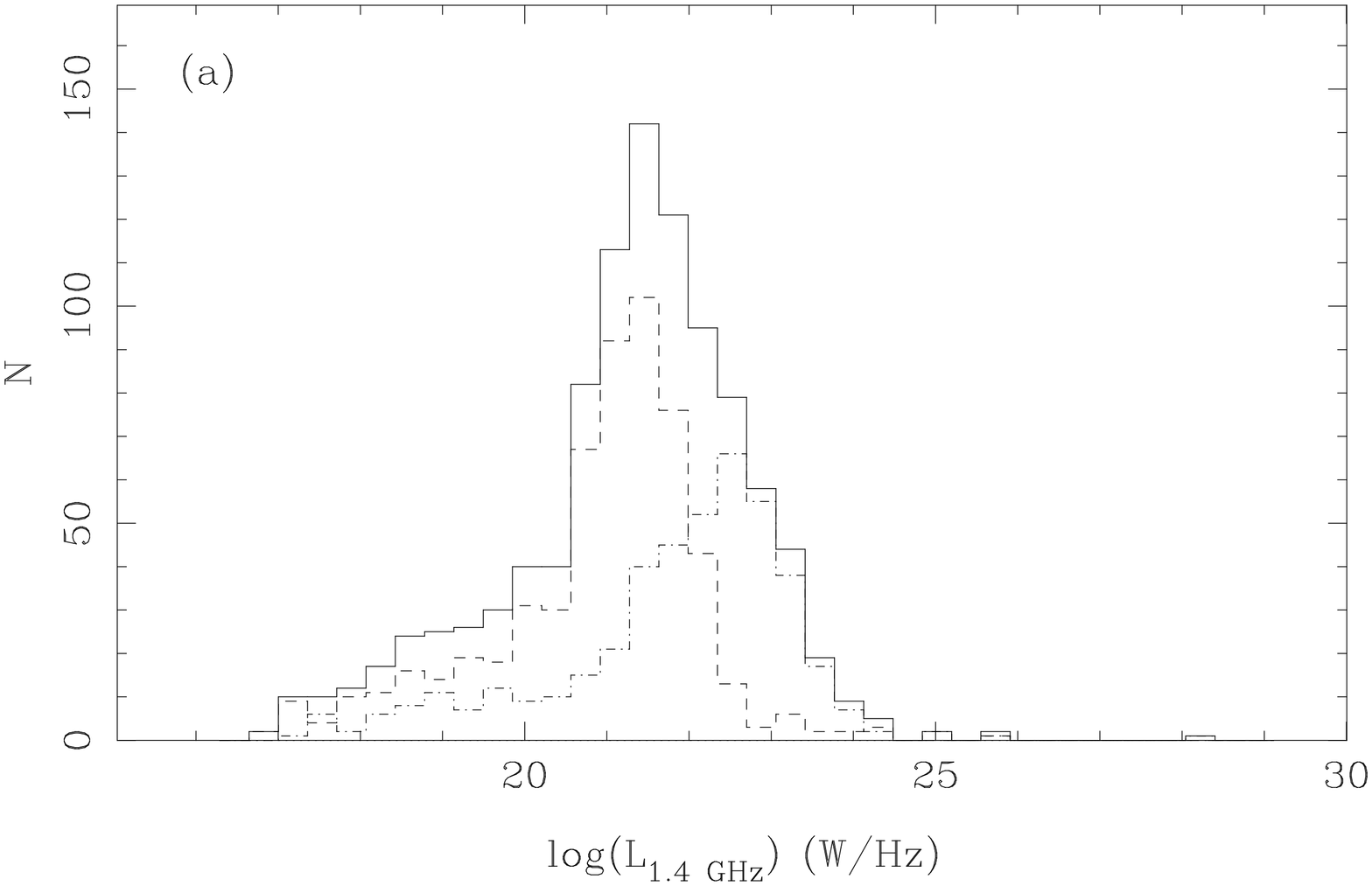,width=5cm,angle=-90}}
\vspace{2mm}
\centerline{\epsfig{figure=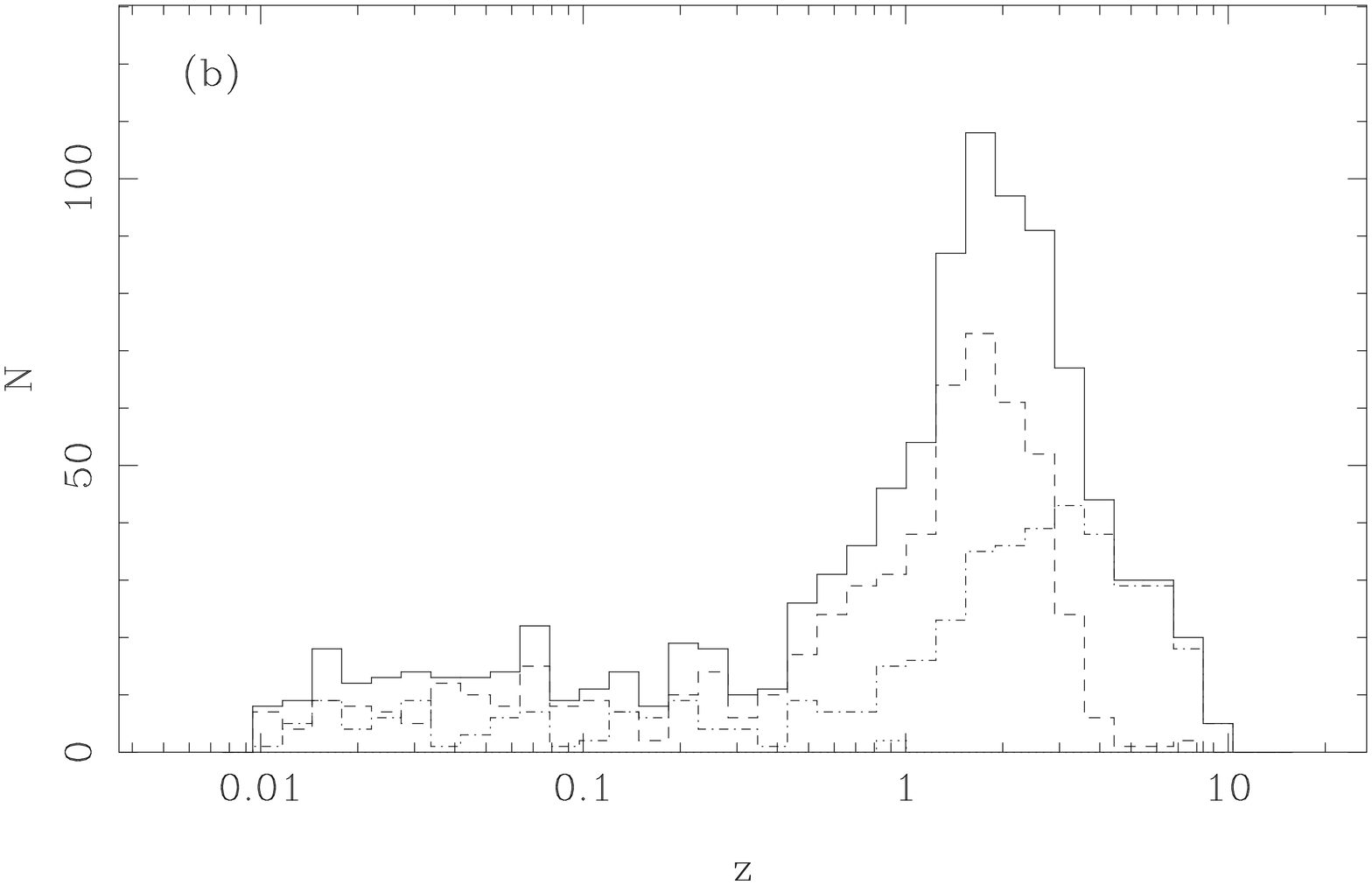,width=5cm,angle=-90}}
\caption{Distribution of (a) luminosities and (b) redshifts
for the simulation in Figure~\protect\ref{pics1}. The line styles correspond
to the different populations. Dashed: AGN; Dot-dashed: IRAS-type; Solid: the
sum of all populations.}
\label{ld1}
\end{figure}

The first example is shown in Figure~\ref{pics1}, a simulation with
a field size of $5'\times5'$, about four times the size of the Hubble Deep
Field (HDF) (c.f. \opencite{Rich:98}). For this simulation the starformers
are represented by the IRAS-type population only. A flux density
limit of $0.1\,\mu$Jy has been used and over 1000 sources brighter than this
limit are predicted (a source density of $\sim4.8\times10^8$\,sr$^{-1}$).
Consistency checks have been performed to confirm that the simulated
distributions in flux density agree with the models used to produce them.
Source counts for the simulation in Figure~\ref{pics1} are shown in
Figure~\ref{sc1} along with the (continuous) source counts derived from the
assumed luminosity functions and evolution. Typical luminosity and redshift
distributions have been derived from the simulations and are presented in
Figure~\ref{ld1}.

\begin{figure}
\centerline{\epsfig{figure=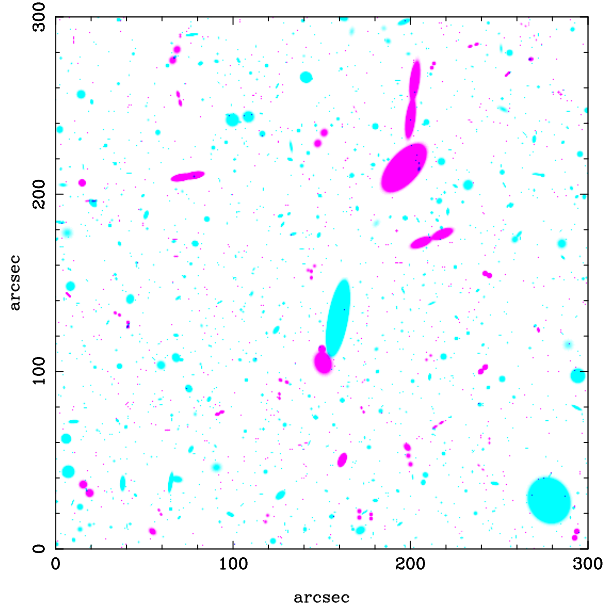,width=8cm,angle=-90}}
\caption{Simulated 1.4\,GHz image. This simulation includes the population
of normal galaxies in addition to the IRAS-types. The image is $5'$ on a
side, with a flux density limit of $0.1\,\mu$Jy. There are 1941 objects in
the simulation, 574 are AGN (red) and 1367 are starformers (blue).}
\label{pics2}
\end{figure}

\begin{figure}
\centerline{\epsfig{figure=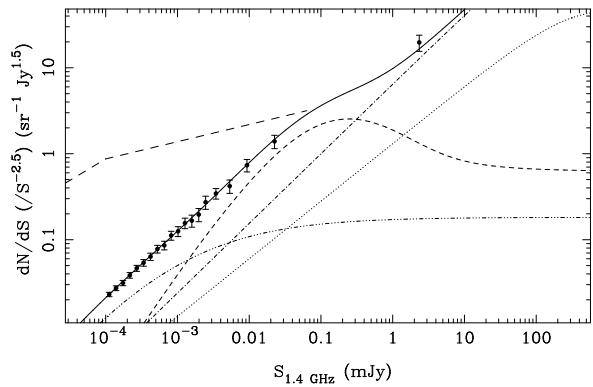,width=5cm,angle=-90}}
\caption{The source counts from the model luminosity functions, along
with those calculated from a typical simulation, to confirm the
self-consistency of the simulations (c.f. Figure~\protect\ref{sc1}). The
new population of normal galaxies (dot-dash-dotted line) begins to make a
noticeable contribution to the source counts below a few microjanskies.
The upper limit to the source counts is again shown as the thick dashed line.}
\label{sc2}
\end{figure}

A second simulation of a $5'\times 5'$ region is shown in
Figure~\ref{pics2}, again with a flux density limit of $0.1\,\mu$Jy,
this time including the population of normal galaxies. A total of
1941 sources (574 AGNs, 1367 starformers) are predicted, giving
a source density of $\sim9.2\times10^8$\,sr$^{-1}$. Again,
source counts, luminosity and redshift distributions are shown
(Figures~\ref{sc2} and \ref{ld2}). It can be seen that even with this scenario
and at these faint levels, while the starformer population is dominant the AGN
population is certainly not negligible. A hint of this effect may be seen in
recent work in the optical domain \cite{Sar:99} showing that $\approx10$\%
of HST Medium Deep Survey galaxies (to $z\approx0.8$) may be low luminosity
AGNs (implied through the presence of unresolved nuclear components that
contribute noticeably to the total galaxy light).
There is also a very clear distinction between the two scenarios (starformers
comprising IRAS-types only compared to the inclusion of a normal galaxy
population) in terms of the number of sources predicted. This is a prediction
that will easily distinguish between the scenarios when SKA
observations are eventually made.

The absence of very large and bright AGN-type sources in the simulated fields
shown is a result of presenting simulated fields coincidentally free of
such objects. This mimics the careful selection against bright sources
typical in deep radio surveys, as well as ensuring clarity in the displayed
simulations. Typically
a simulated field of this size may contain one or two sources of
several tens of millijanskies, and with physical sizes dominating the field.

The limiting flux density value of $0.1\,\mu$Jy was chosen for a variety
of reasons. Primarily, the consistency of our simulations
with existing observations (to levels of $50\,\mu$Jy) implies a reasonable
reliability to levels of about $1\,\mu$Jy, so a level at least an order of
magnitude fainter is necessary to extend the simulations into a new regime,
where the dominant population is truly unknown, and where the actual
properties of the high-redshift universe may be revealed by future
observations. Also, for an interferometer having the specifications given for
the SKA, $0.1\,\mu$Jy is the expected $5\sigma$ level achievable after a 12
hour integration. The relevant specifications are an effective area to system
temperature ratio of $A_{\rm eff}/T_{\rm sys}=2\times10^4$\,m$^2$K$^{-1}$
with two simultaneous frequency bands and a bandwidth $\Delta\nu=500\,$MHz at
1.4\,GHz \cite{BT:99}.
Images deeper than this $0.1\,\mu$Jy level will obviously be attainable through
longer integrations with the instrument, (although predictions to such levels
are likely to suffer from increasing uncertainties in the extrapolations).
For example, the time spent observing the HDF with the Very Large Array
\cite{Rich:98} was 152 hours. Some recent observations of
similar integration time with the Australia Telescope Compact Array are
provided by \inlinecite{Hop:99b} and \inlinecite{Nor:99}. With the
potential to reach a $5\sigma$ detection level of about 30\,nJy over such an
integration time, dynamic range considerations for the SKA
will be very important! In the HDF observations, the VLA
was dynamic range limited at a level of 10,000:1. The ATCA deep field
observations of the HDF-S are not yet dynamic range limited, but have a
thermal noise limit of 10,000:1. A dynamic range of
100,000:1 has been reached by ATCA in other wide field images.
Although no insurmountable technical difficulties are anticipated in
reaching the required levels of $10^6$:1 or $10^7$:1 on axis,
achieving similar values at the half-power level will be more difficult,
and this is an aspect of the project which deserves close attention
(see also \opencite{deB:96}).

\section{Discussion}

\begin{figure}
\centerline{\epsfig{figure=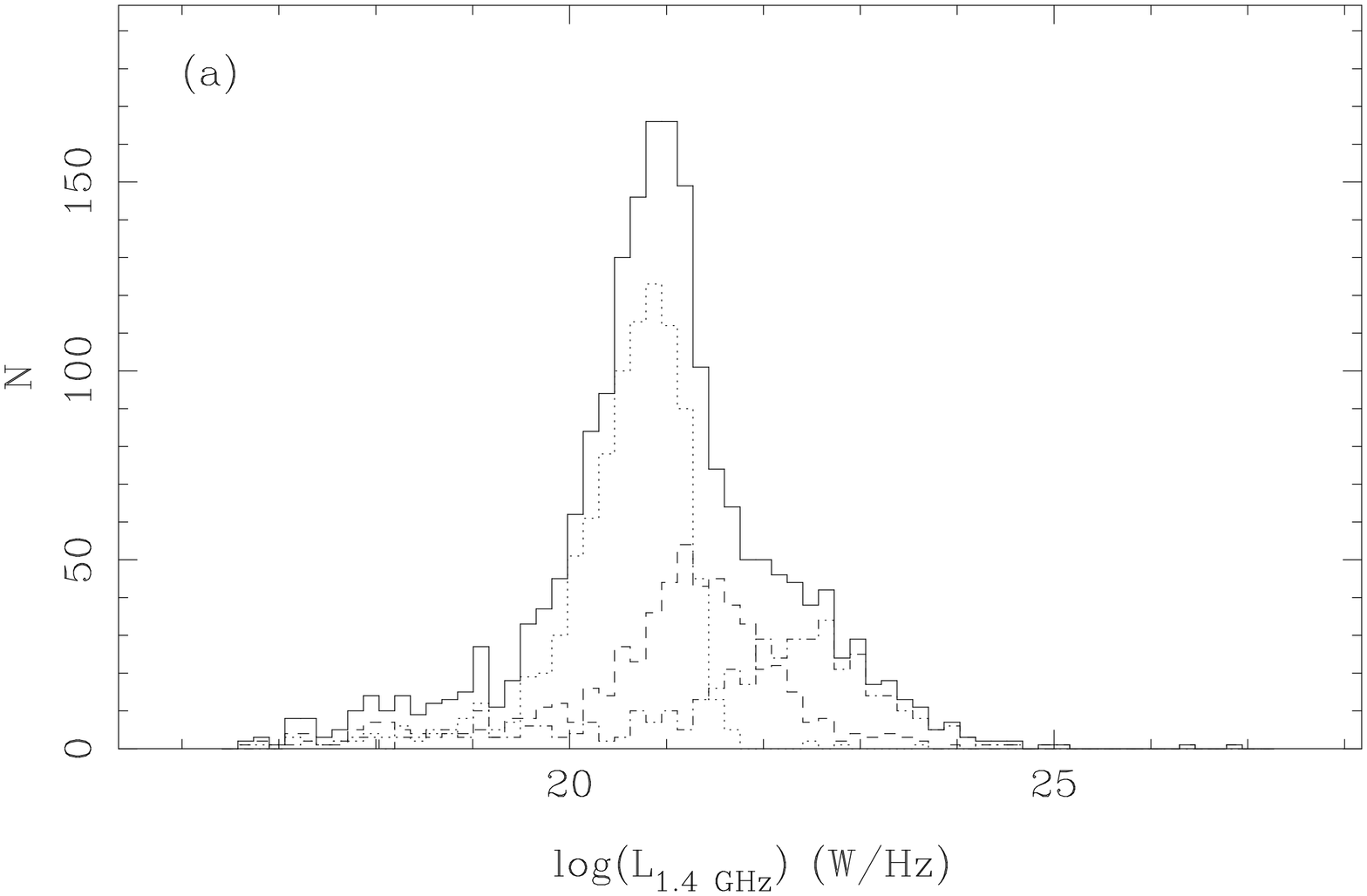,width=5cm,angle=-90}}
\vspace{2mm}
\centerline{\epsfig{figure=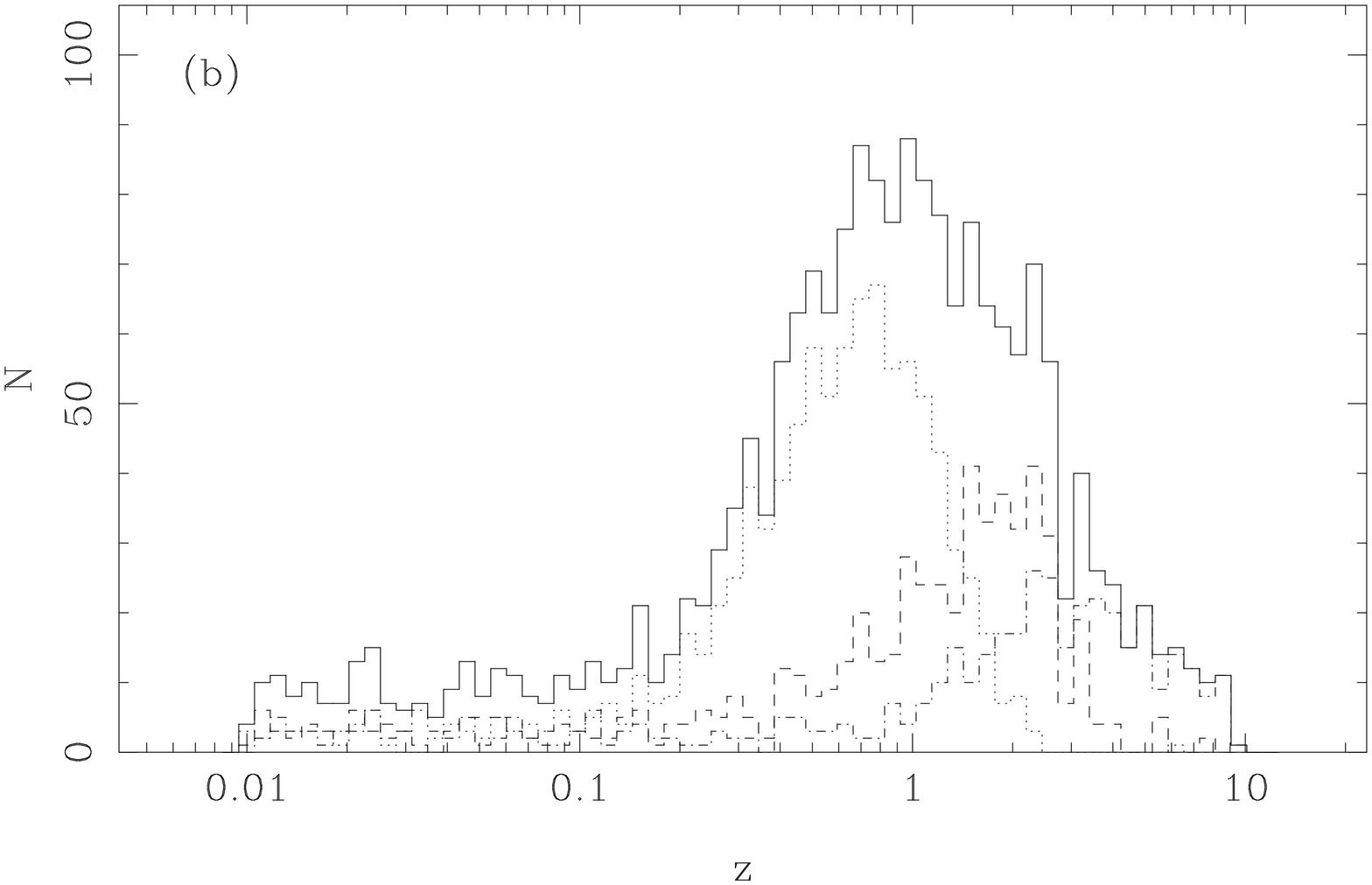,width=5cm,angle=-90}}
\caption{Distributions of (a) luminosity and (b) redshift
for simulation in Figure~\protect\ref{pics2}. The line styles correspond to
the different populations. Dashed: AGN; Dot-dashed: IRAS-type; Dotted: normal
galaxies; Solid: the sum of all populations.}
\label{ld2}
\end{figure}

In each simulation shown, the brightest source is an AGN, but between
luminosities of about $10^{22}\,$WHz$^{-1}$ and $10^{24.5}\,$WHz$^{-1}$ the
IRAS-types dominate (see Figures~\ref{ld1} and \ref{ld2}).
This result can be explained as a combination of the small area being sampled
(as the bright AGNs have a low surface density), and the dearth of AGN-type
galaxies further than $z \approx 3$. This redshift cutoff in the distribution
of AGNs results from the chosen model of luminosity evolution
for this population \cite{DP:90}. The redshift distribution for the
``normal" galaxies (Figure~\ref{ld2}) on the other hand, which shows no
objects further than redshifts of $z \approx 2$ in these simulations, is not
a result of the models used but is instead dependent on the
flux density limits chosen for the simulation. Had the limits been fainter,
such objects (having only modest intrinsic luminosities, predominantly
$< 10^{22}\,$WHz$^{-1}$, as can be seen from the luminosity distribution
in Figure~\ref{ld2}) would have been detected at continually higher
redshifts. Additionally, the distribution of luminosities for this population
is very narrow, concentrated primarily within two orders of magnitude.
This reflects the shape of the optical luminosity function from 
which this population is derived, and it is useful to emphasise the very
small range of intrinsic luminosities displayed by galaxies at optical
wavelengths compared to the range of ten orders of magnitude or more possible
in a galaxy's intrinsic radio luminosity.

\begin{table}[h]
\caption{Specifications necessary for a radio telescope able to image a
one square degree field of view to a limiting $5\sigma$ flux density of
$S_{\rm min}$ in a 12\,hr integration.}
\label{spectab}
\begin{tabular}{cccccc}
\hline
$S_{\rm min}$ & N & DR & $A_{\rm eff}$/$T_{\rm sys}$ & $\theta_{\rm min}$ & $B_{\rm max}$ \\
($\mu$Jy) & & & (m$^2$K$^{-1}$) & ($''$) & (km) \\
\hline
1.00 & 360  & $10^5$ & $2\times10^3$ & 1.5 & 30 \\
0.10 & 1941 & $10^6$ & $2\times10^4$ & 0.6 & 70 \\
0.01 & 9431 & $10^7$ & $2\times10^5$ & 0.3 & 150 \\
\hline
\end{tabular}
\end{table}
Table~\ref{spectab} shows how the number of sources increases as the limiting
($5\sigma$) flux density is reduced, in the scenario where starformers include
both IRAS-types and normal galaxies. Column 3 gives the dynamic range necessary
if there will regularly be 100\,mJy sources in the (1 square degree) field of
view. Very deep radio surveys will obviously continue to be carried out in
carefully selected regions, chosen to be free of bright sources. But the
specified field of view of 1 square degree at 1.4\,GHz implies all but a few
areas chosen for deep observations are likely to contain sources of at least
several tens of millijanskies. The effective area to system temperature ratio
for an instrument which can reach the given limiting flux density in a 12 hour
integration is given in column 4 of Table~\ref{spectab}. Columns 5 and 6
show the minimum resolution and corresponding maximum baseline for operation
at 1.4\,GHz required to avoid formal (instrumental) confusion, corresponding
simply to the number of sources regardless of their apparent sizes.
The specification $A_{\rm eff}/T_{\rm sys}=2\times10^4\,$m$^2$K$^{-1}$
corresponds to an effective collecting area of 1 square kilometre with a
system temperature of 50\,K. This configuration could be used (instead of those
in rows 1 and 3 of the Table) to reach the limiting flux densities of
$1\,\mu$Jy and $0.01\,\mu$Jy by integrating for 7 minutes and 1200 hours,
respectively. While a resolution of $0\farcs6$ is required to avoid formal
confusion at the $0.1\,\mu$Jy level, this would drop to $0\farcs4$ (and lower
for fainter flux density limits) if the source count achieves the maximum
level given the CBR limitations. This could arise through new populations not
included in the simulations presented here, or different evolutionary forms
or rates to those presented in these simulations.

While the limited overlapping of objects in the simulated images implies that
the natural confusion level has not yet been reached for the majority of
sources, and only the largest and brightest are likely to be affected,
a high resolution ($\sim0\farcs1$) is desirable to adequately
resolve superimposed sources, facilitating their study.
The extent of the natural confusion is directly related to the intrinsic
distribution of linear sizes for the starforming population (as opposed to the
models used in these simulations). If the true distribution is dominated by
objects with small half-light radii (only a few kpc, c.f. \opencite{Ode:96})
the effects of natural confusion, to the $0.1\,\mu$Jy level
investigated here, will be mild. If the source count reaches its potential
limits, though, the problem could be more significant.
For high imaging quality at the faintest flux densities (and smallest angular
scales) the SKA will require a high sensitivity
response over the longer baselines. This implies that a significant fraction
of the collecting area needs to be distributed over baselines up to hundreds
of kilometres for operation at 1.4\,GHz ($0\farcs1$ resolution requiring
baselines of $\sim440\,$km). There will also be a conflict between designs
optimised for operation at high resolution and for good surface brightness
sensitivity, so an optimum solution that best addresses the desires of the
astronomical community will need to be established. If the instrument is built
with a good response at high resolution, then the effects of confusion will be
minimal, and very deep radio surveys with the SKA
are much more likely to be dynamic range limited than confusion limited.

The large field of view specified for this instrument is 144 times the area of
the simulations presented here, hence observations down to $0.1\,\mu$Jy
will result in $\sim3\times10^5$ sources per field.
This has implications for the nature and sophistication of automated source 
detection and characterisation algorithms that must be developed to analyse
such data. Also, since a well sampled image will have several pixels across
a synthesised beam FWHM (say $0\farcs1$), an image of a one
square degree field will be $\sim10^5\times10^5$ pixels in size.
Although the tens of gigabytes per image required for storage will not
be a problem, visual inspection of entire images at full resolution
will become difficult in the extreme, and will likely be reserved for objects
denoted as interesting by automated classifiers.

\section{Conclusions}

We have initiated a program of simulating the 1.4\,GHz sky to investigate
the implications of existing models for luminosity functions, luminosity
evolution and linear size distributions on anticipated observations by the SKA.
The results are self-consistent and consistent
with observed source counts, although below that level it should be remembered
that they are based on {\em extrapolations}. The corresponding level of
uncertainty at ever fainter flux densities will increase, and in
fact we might expect to find something quite new and different when
actual observations can be made at these sensitivities.
Many simulations have been investigated and two of them
presented here. The simulations produced with a flux density limit of
$0.1\,\mu$Jy show fields that are dominated by starforming galaxies, although
the proportion of AGN galaxies is still significant, even at such faint flux
densities. The level of natural confusion in these fields is very strongly
related to the choice of models for intrinsic linear size of the starformers,
and deserves closer attention. To model different rates of evolution and
different cosmologies, additional luminosity function models
that correctly reproduce the observed source counts in each of the chosen
scenarios are necessary.

The preliminary specifications for the SKA
\cite{BT:99} have been discussed in the context of the simulations presented
here. There would appear to be only minor problems due to natural confusion
at the levels of $0.1\,\mu$Jy, and no problems with instrumental confusion,
even at much lower flux densities, if a resolution around
$0\farcs1$ to $0\farcs3$ is chosen.
The large field of view and high resolution will necessarily
affect the way images are analysed, reducing the level of visual
interpretation.

\begin{acknowledgements}
We wish to thank the referee for useful suggestions.
AMH and LEC acknowledge financial support from the Australian Research
Council and the Science Foundation for Physics within the University of
Sydney. 
RAW acknowledges grant AST-9802963 from the National Science Foundation,
and NASA grants GO-5985.01-94A and GO-6609.01-95A from STScI under NASA
contract NAS5-26555.
The Australia Telescope is funded by the Commonwealth of Australia for
operation as a National Facility managed by CSIRO.
\end{acknowledgements}

\clearpage
 
\end{article}
\end{document}